\documentclass[fleqn,twoside]{article} 
\usepackage{epsf,multicol,ifthen}
\usepackage{ujp}
\usepackage[cp1251]{inputenc}
\usepackage[english,russian]{babel}
\usepackage{latexsym, amssymb}

\nazva{APPLYING THE \boldmath$q$-ALGEBRAS \boldmath$U'_q({\rm
so}_n)$ TO QUANTUM GRAVITY:
TOWARDS \boldmath$q$-DEFORMED ANALOG\\ OF \boldmath${\rm SO}(n)$ SPIN NETWORKS\footnotemark[1]}%

\udk{538.9; 538.915; 517.957}

\nazvacol{APPLYING THE $q$-ALGEBRAS $U'_q({\rm so}_n)$ TO QUANTUM GRAVITY}%

\avtor{A. M. GAVRILIK}%
\avtorcol{A. M. GAVRILIK}%
\inst{Bogolyubov Institute for Theoretical Physics, Nat. Acad. Sci. of Ukraine}%
\adr{(14b, Metrolohichna Str., 03143 Kyiv, Ukraine)}%

\begin{document}           
\setcounter{page}{213}%

\newcommand {\be}{\begin{equation}}
\newcommand {\ee}{\end{equation}}
\newcommand{\bea}{\begin{eqnarray}}
\newcommand{\eea}{\end{eqnarray}}
\newcommand{\ba}{\begin{array}}
\newcommand{\ea}{\end{array}}

\newcommand{\Uqso}{{U'_q({\rm so}_n)}}


\maketitl                 

\begin{multicols}{2}
\anot{%
Nonstandard $q$-deformed algebras $U'_q({\rm so}_n)$, proposed a
decade ago for the needs of representation theory, essentially
differ from the standard Drinfeld---Jimbo quantum deformation of
the algebras $U({\rm so}_n)$ and possess with regard to the latter
a number of important advantages. We discuss possible application
of the $q$-algebras $U'_q({\rm so}_n)$, within two different
contexts of quantum/$q$-deformed gravity: one concerns
$q$-deforming of $D$-dimensional ($D\ge 3$) euclidean gravity, the
other applies to 2+1 anti-de Sitter quantum gravity (with space
surface of genus $g$) in the approach of Nelson and
Regge.\footnotetext[1]{Presented at the XIIIth International
Hutsulian Workshop ``Methods of Theoretical and Mathematical
Physics'' (September 11 --- 24 2000, Uzhgorod --- Kyiv ---
Ivano-Frankivsk
--- Rakhiv, Ukraine) and dedicated to Prof. Dr. W.~Kummer on the
occasion of his 65th birthday.}
}%

\section{Introduction}

\noindent Construction of quantum gravity belongs to most
fundamental problems of modern quantum theory. During last decade
and a half, new perspective tools for attacking and solving this
problem have appeared among which we mention, first, the notion of
spin networks (see e.g., \cite{RoSmo}) closely connected with loop
quantum gravity as well as with the so-called BF-type topological
theories, and, second, the powerful methods of quantum groups and
quantum algebras [2 --- 5]. Our goal in this contribution is to
consider potential applicability of the so-called nonstandard
$q$-deformed algebras $U'_q({\rm so}_n)$ introduced in \cite{GK91}
which are different from the standard (Drinfeld---Jimbo) quantum
deformation \cite{Dri,Ji} of the Lie algebras of orthogonal
groups, while possess a number of rather important advantages.
Here we intend to make a preliminary steps towards extending the
$D$-dimensional version of spin networks (more concretely, ${\rm
SO}(D)$ simple spin networks) to the case of $U'_q({\rm so}_n)$
related formulation. In the second part of our contribution, we
briefly discuss the appearance of the $U'_q({\rm so}_n)$ algebras
in the context of anti-de Sitter 2+1 quantum gravity formulated
with space-part being fixed as genus $g$ Riemann surface so that
$n=2g+2$.



\section{ Simple \boldmath$G={\rm SO}(n)$ Spin Networks}


Let us first briefly dwell upon necessary setup concerning $G={\rm
SO}(n)$ spin networks.

A generalized spin network associated with
a Lie group $G$, according e.g. to                   \cite{FrKr},
is defined as a triple $(\Gamma, \rho, I)$ where

$\Gamma$ is an oriented graph, formed by directed edges and vertices;

$\rho$ is a labeling of each edge $e$ by an irreducible representation
(irrep) $\rho_e$ of $G$;

$I$ is a labeling of each vertex $v$ of $\Gamma$ by an intertwinner
$I_v$ mapping tensor product of irreps incoming at $v$ to the
product of irreps outgoing from $v$.

Below, we are interested in the spin networks for the particular
Lie group $G={\rm SO}(n)$. Moreover, like in \cite{FrKr}, we
consider restricted case of $G={\rm SO}(n)$ {\em simple} spin
networks. Simple spin networks associated with $G={\rm SO}(n)$ are
evaluated as Feynman integrals over the coset space ${\rm
SO}(n)/{\rm SO}(n-1)$, i.e. over the sphere $S^{n-1}$. Simplicity
means that only the ${\rm SO}(n)$ representations of class 1 (with
respect to ${\rm SO}(n-1)$) labeled by single nonnegative integer
$l$, are employed.

Basic ingredient is the `propagator' expressed in terms
of zonal spherical functions
$t^{nl}_{00}(y), \ y=\cos\theta$, or, in view of the equality   \cite{VK}
\be
t^{Nl}_{00}(\cos\theta)=
\frac{\Gamma(2p) l!}{\Gamma(2p+l)} C^p_l(\cos\theta)\ , \ \ \ \ p=(N-2)/2\ ,
\ee
directly through
the Gegenbauer polynomials:
\be
G_m^{(N)}(x,y)= \frac{N+2m-2}{N-2}C_m^{(N-2)/2}(x{\cdot}y)\ .
\ee
Here the Gegenbauer polynomials $C^p_m(x)\ ,$ $\ l\ge 0$, 
satisfy  
the defining recursion relation
\[
(l+1)C^p_{l+1}(x)=
\]
\be
2(p+l) x C^p_{l}(x) - (2p+l-1)C^p_{l-1}(x)
\ee
augmented with the initial value $C^p_{0}(x)=1$, and obey
the orthogonality relation
\[
\int^1_{-1}(l+1)C^p_{l}(x) C^p_{m}(x) (1-x^2)^{p-\frac12}{\rm d}x =
\]
\be
=\delta_{lm}\cdot\frac{\pi \Gamma (2p+l)}{2^{2p-1} l! (l+p) \Gamma^2(p)}\ .
\ee
Another important property is given by the linearization formula
for the product of two Gegenbauer polynomials, namely
\[
C^p_{l}(x) C^p_{m}(x) = \sum^{l+m}_{n=\vert l-m\vert }
\frac{(n+p)\Gamma(n+1)\Gamma(g+2p)}{[\Gamma(p)]^2 (g+p)!
\Gamma(n+2p)}\times
\]
\be
 \times \frac{\Gamma(g-l+p)\Gamma(g-m+p)\Gamma(g-n+p)}
       {\Gamma(g-l+1)\Gamma(g-m+1)\Gamma(g-n+1)}\ C^p_{n}(x)
\ee
where $g\equiv\frac12 (l+m+n)$ and the sum ranges over those values
of $n$ which are of the same evenness as $l+m+n$.

One of basic constructs 
in spin networks is the so-called $\Theta$-graph whose evaluation
is given by the expression
\be
D(l,m,n;p):=\!\int_{-1}^1\!C^p_{l}(x)
C^p_{m}(x) C^p_{n}(x)(1-x^2)^{p-\frac12}{\rm d}x . \
\ee
The result of evaluation is
\[
D(l,m,n;p)= \frac{2^{1-2p} \pi}{[\Gamma(p)]^4}
       \frac{\Gamma(g+2p)}{\Gamma(g+p+1)} \times
\]
\be
 \times \frac{\Gamma(g-m+p)\Gamma(g-n+p)}
           {\Gamma(g-l+1)\Gamma(g-m+1)\Gamma(g-n+1)}\ .
\ee
Using (6) one easily deduces the recurrence relation for
the $D(l,m,n;p)$ in the form
\[
\frac{l+1}{p+l} D(l+1,m,s;p)+\frac{2p+l-1}{p+l} D(l-1,m,s;p) =
\]
\be
=
\frac{s+1}{p+s} D(l,m,s+1;p)+\frac{2p+s\!-\!1}{p+s} D(l,m,s\!-\!1;p) \
\ee
(remark that in 4 dimensions all the multipliers become equal to 1).

Thus, $\Theta$-graph $\theta^{(N)}(m_1,m_2,m_3)$ is nothing but
integral of the product of three  
normalized propagators defined in (2):
\[
\theta^{(N)}(m_1,m_2,m_3)=
\]
\be
= \frac{\Gamma(g+2p)\Gamma(p+1)}{\Gamma(g+p+1)\Gamma(2p)}
\prod^3_{i=1}
\frac{(m_i+p)\Gamma(g-m_i+p)}{\Gamma(p+1)\Gamma(g-m_i+1)},
\ee
where $g=(m_1+m_2+m_3)/2$ is an integer, $g-m_i\ge 0$, $i=1,2,3,$ and
$p=(N-2)/2$.

For $N=4$,
\be
\theta^{(4)}(m_1,m_2,m_3)= (m_1+1)(m_2+1)(m_3+1).
\ee
There exist a number of other results concerning (simple) ${\rm
SO}(n)$ spin networks, for generic situation as well as for the
particular cases of $n=3,4$, which we however shall not discuss
here further.

\section{\boldmath$q$-Deformed Analog of Spin Networks from
\boldmath$q$-ultraspherical Polynomials}

To deal with $q$-deformed case we need some facts concerning
$q$-ultraspherical polynomials.
These are defined through the following recursion relations:
\[
(1-q^n)C_n(x;\beta|q)=2x(1-\beta q^{n-1}) C_{n-1}(x;\beta|q)
\]
\be
-(1-\beta^2 q^{n-2}) C_{n-2}(x;\beta|q), \ \ \ \ \ \ (n\ge 2),
\ee
along with special values
\be
C_0(x;\beta|q)=1, \ \ \ \ C_1(x;\beta|q)=2 (1-\beta) x /(1-q).
\ee
With $\beta = q^{\lambda}$, the "classical" limit $q\to 1$ yields
\be
C_n(x;\beta|q)
       \stackrel{ q\to 1}{-\!\!\!\longrightarrow } C_n^{\lambda} (x).
\ee
The explicit expression for the $q$-ultraspherical polynomials is  \cite{GR}
as follows:
\[
C_n(x;\beta|q) = \sum^n_{k=0}
 \frac{(\beta;q)_k (\beta;q)_{n-k}}
      {(q;q)_k (q;q)_{n-k}}
      {\rm e}^{{\rm i}(n-2k)\theta}
\]
\be
= \frac{(\beta;q)_n}{(q;q)_n} {\rm e}^{in\theta}
{}_2\Phi_1 (q^{-n},\beta; \beta^{-1}q^{1-n}; q,q\beta^{-1}{\rm e}^{-2i\theta}) .
\ee
In this formula, the notation $(a;q)_n$ means:
\be
(a;q)_n = \cases{\ \ 1, \ \ \ \ \ \ n=0 \cr
            (1-a)(1-qa)...(1-q^{n-1}a),\ \ \ \ \ n\ge 1 . \cr
}
\ee
It should be noted that it is also possible to present
$C_n(x;\beta|q)$ in the form which employs basic
hypergeometric function ${}_4\Phi_3$ or ${}_3\Phi_2$, see    \cite{GR,AW}.

Orthogonality relations for the $q$-ultraspherical polynomials
are of principal importance. They are given by the relation
\[
\int\limits_0^{\pi}
C_m(\cos\theta;\beta|q) C_n(\cos\theta;\beta|q)
W_{\beta}(\cos\theta|q){\rm d}\theta
=
\]
\vspace{-3mm}
\be
= \frac{\delta_{mn}}{h_n(\beta|q)}, \ee where the weight function
and the normalization factor are as follows:
\be
W_{\beta}(\cos\theta|q)
=\frac{({\rm e}^{2i\theta},{\rm e}^{-2i\theta};q)_{\infty}}
      {(\beta{\rm e}^{2i\theta},\beta{\rm e}^{-2i\theta};q)_{\infty}} ,
\ee
\be
h_n(\beta |q)
=\frac{(q,\beta^2;q)_{\infty}(q;q)_n(1-\beta q^n)}
      {2\pi (\beta,\beta q;q)_{\infty}(\beta^2;q)_n(1-\beta )} ,
\ee
with
\[  (a_1,a_2;q)_\infty := (a_1;q)_\infty (a_2;q)_\infty \ ,
\]
\[              
     (a;q)_{\infty}:= \prod_{k=0}^{\infty} (1-aq^k) .
\]
%
Linearization formula is another important fact about
$q$-ultraspherical polynomials.
It is given by the following Rogers' formula         \cite{GR}:
\[
C_m(x;\beta|q) C_n(x;\beta|q)=
\]
\be
=\sum_{k=0}^{\min(m,n)} A_{m,n,k}(\beta|q) C_{m+n-2k}(x;\beta|q)
\ee with the notation
\[
A_{m,n,k}(\beta|q) =
\frac{ (\beta;q)_{m-k}\ (\beta;q)_{n-k}\ (\beta;q)_k}
     { (q;q)_{m-k}\ (q;q)_{n-k}\ (q;q)_k             }  \times
\]
\be
 \times \frac{ (q;q)_{m+n-2k}\ (\beta^2;q)_{m+n-k}  }
         {(\beta^2;q)_{m+n-2k}\ (\beta q;q)_{m+n-k} } \ 
         \frac{\ \ \ 1-\beta q^{m+n-2k} }{1-\beta }.
\ee
\medskip
Now it is not hard to obtain the result for the $q$-deformed analog of
$\Theta$-graph (9), namely   
\[
D_q(m,n,s,\lambda) =
\]
\[   \vspace{-4mm}
 = \int\limits_0^{\pi}
C_m(x;\beta|q) C_n(x;\beta|q) C_s(x;\beta|q)
W_{\beta}(\cos\theta|q){\rm d}\theta =
     \vspace{-4mm}
\]
\be
=\frac{ 2\pi \ (\beta,\beta q;q)_{\infty} (\beta^2;q)_g
             (\beta;q)_{g-n}(\beta;q)_{g-m}(\beta;q)_{g-s}    }
      { \ \ \ (q,\beta^2;q)_{\infty} (\beta q;q)_g
                (q;q)_{g-n} (q;q)_{g-m} (q;q)_{g-s}           } ,
\ee
where $m+n+s=2g$, $\ g\ge m$, $g\ge n$, $g\ge s$.
   Notice the obvious symmetry under exchanges:
$m\leftrightarrow n\leftrightarrow s\leftrightarrow m$.
Recursion relation for $D_q$ is obtained in the form 
\[
\frac{1-q^{m+1}}{1-\beta q^m }  D_q(m+1,n,s,\lambda) +
\]
\[
+ \frac{1-\beta^2 q^{m-1}}{1-\beta q^m }  D_q(m-1,n,s,\lambda) =
\]
\[
= \frac{1-q^{s+1}}{1-\beta q^s }   D_q(m,n,s+1,\lambda) +
\]
\be
+ \frac{1-\beta^2 q^{s-1}}{1-\beta q^s }   D_q(m,n,s-1,\lambda) .
\ee
Likewise, one can get evaluation for other particular
($q$-deformed analogs of) spin networks.

\section{Covariance with Respect to \boldmath$q$-algebras}

Our main concern here is a possible relation of this stuff to
quantum groups and/or $q$-algebras which correspond to the
orthogonal Lie groups ${\rm SO}(n)$ and their corresponding Lie
algebras. As it was shown by Sugitani in  \cite{Sugi}, zonal
spherical functions associated with a particular realization
$S^N_q$ of quantum spheres are proportional to the
$q$-ultrapsherical polynomials, that is \vskip2mm
\noindent($q$-)zonal spher. func. $\leftrightarrow
C_k^{(N-2)/2}(Y; q^2)$ . \vskip2mm \noindent To this end, one
starts with standard $U_q({\rm so}_n)$ and constructs a $q$-analog
of the coset ${\rm SO}(n)/{\rm SO}(n-1)$ by means of $U_q({\rm
so}_n)$ (corresponding to ${\rm SO}(n)$) and a coideal
(corresponding to ${\rm SO}(n-1)$):
\be
J_q:=\cases{
\left\langle e_2,...,e_n,f_2,...,f_n,\theta_1,\theta_2,
\frac{q^{\epsilon_2} -1}{q-1},...,\frac{q^{\epsilon_n} -1}{q-1}
                           \right\rangle ,          \cr
                             {}                    \cr
       \hspace{2mm} {\rm for}\ B_n (n>1)\
    {\rm and} \ \  D_n (n>2) \ \ {\rm series} ,   \cr
                             {}                    \cr
        \langle \theta_1 \rangle \ \ \ \ {\rm for}\ N=3       \cr
                              {}                    \cr
\langle\theta_1, \theta_2, \frac{q^{\epsilon_2} -1}{q-1}\rangle
\ \ \ {\rm for}\ \ N=4\ . \cr
       }
\ee Here \[ \theta_1 :=\cases{
  s\cdot e_1 + (-1)^{n-1} t\cdot q^{1/2} q^{\epsilon_1}
                        f_2\cdots f_n f_n \cdots f_2 f_1    \cr
                                 \vspace{-3mm}      \cr
       \hspace{2mm} {\rm for}\ B_n\ (n>1)\ \ {\rm series} ,  \cr
                                  \vspace{-0.1mm}      \cr
 s\cdot e_1 + (-1)^{n-2} t\cdot q^{\epsilon_1}
                f_2\cdots f_{n-1} f_n f_{n-2} \cdots f_2 f_1  \cr
                                  \vspace{-3mm}      \cr
       \hspace{2mm} {\rm for}\ D_n\ (n>2)\ \ {\rm series} ,    \cr
                                   \vspace{-0.2mm}          \cr
s\cdot e_1 + t\cdot q^{1/2} q^{\epsilon_1} f_1  \ \ \ (N=3) ,   \cr
                                   \vspace{-0.3mm}           \cr
s\cdot e_1 + t\cdot q^{\epsilon_1} f_2  \ \ \ (N=4) ;   \cr
                }
\]\begin{equation}\end{equation}
\[
\theta_2 :=\cases{
  t\cdot q^{1/2} q^{\epsilon_1} f_1 +
   (-1)^{n-1} s\cdot e_2\cdots e_n e_n \cdots e_2 e_1    \cr
                                    \vspace{-3mm}      \cr
      \hspace{2mm} {\rm for}\ B_n\ (n>1)\ \ {\rm series} , \cr
                                    {}               \cr
 t\cdot q^{\epsilon_1} f_1 + (-1)^{n-2}
       s\cdot e_2\cdots e_{n-1} e_n e_{n-2} \cdots e_2 e_1   \cr
                                    \vspace{-3mm}      \cr
       \hspace{2mm} {\rm for}\ D_n\ (n>2)\ \ {\rm series} ,  \cr
                                    \vspace{-0.3mm}      \cr
       t\cdot q^{\epsilon_1} f_1 + s\cdot e_2 \ \ \ (N=4) ,   \cr
                }
\]\begin{equation}\end{equation}
As follows from the results of \cite{Sugi}, this left coideal
subalgebra coincides with the nonstandard $q$-deformed algebra
$U'_q({\rm so}_n)$ from \cite{GK91}. Note also that this same
nonstandard (or twisted) $q$-deformed coideal subalgebra arises
\cite{Nou,NUW96} when one constructs a quantum analogue of the
symmetric coset space SU$(n)/{\rm SO}(n)$.

\section{ The \boldmath$q$-algebra \boldmath$U'_q({\rm so}_n)$ (Bilinear Formulation)}

Along with the definition in terms of trilinear relations
originally given in           \cite{GK91}, the $q$-algebra
$U'_q({\rm so}_n)$ may be equivalently defined in terms of {\it
`bilinear' formulation}. To this end, the generators (set $k >
l+1, \ \  1\leq k,l \leq n$)
\[
I^{\pm}_{k,l}\equiv
[I_{l+1,l} , I^{\pm}_{k,l+1}]_{q^{\pm 1}}
\equiv
\]
\[
\equiv
q^{\pm 1/2}I_{l+1,l} I^\pm_{k,l+1} -
q^{\mp 1/2}I^\pm_{k,l+1} I_{l+1,l}
\]
are introduced together with $I_{k+1,k} \equiv I^+_{k+1,k}\equiv
I^-_{k+1,k}$. Then, the bilinear formulation of the $q$-algebra
$U'_q({\rm so}_n)$ reads:
\[
[I^+_{lm} , I^+_{kl}]_q = I^+_{km}, \ \
[I^+_{kl} , I^+_{km}]_q = I^+_{lm},
\]
\[
[I^+_{km} , I^+_{lm}]_q = I^+_{kl}  \ \ {\rm if} \  k>l>m,
\]
\be                                               \label{f2}
[I^+_{kl} , I^+_{mp}] = 0 \ \  {\rm if} \   k\!>\!l\!>\!m\!>\!p
\ \ \ {\rm or } \ \  k\!>\!m\!>\!p\!>\!l;
\ee
\[
[I^+_{kl} , I^+_{mp}] =
(q\!-\!q^{-1}) (I^+_{lp}I^+_{km}\!-\!I^+_{kp}I^+_{ml})
 \ \ {\rm if}  \ k\!>\!m\!>\!l\!>\!p.
\]
Analogous set of relations exists which involves $I_{kl}^-$ along with
$q\to q^{-1}$ (denote this ``dual'' set by (\ref{f2}$'$)).
In the `classical' limit $q\to 1$ , both (\ref{f2}) and
(\ref{f2}$'$) reduce to those of ${\rm so}_n$.

For instance, at $n=3$,
the $q$-algebra $U'_q({\rm so}_3)$ is isomorphic       \cite{GK94}
to the Fairlie -- Odesskii algebra                    \cite{Od,Far}
(recall that the $q$-commutator is defined as
$[X,Y]_q\equiv q^{1/2}X Y - q^{-1/2} Y X$):
\be
[I_{21} , I_{32}]_q = I_{31}^+, \
[I_{32} , I_{31}^+]_q = I_{21}, \
[I_{31}^+ , I_{21}]_q = I_{32}; \                     \label{FO}
\ee
at $n=4$ the $q$-algebra $U'_q({\rm so}_4)$ in addition involves:
\[
\hspace{-1.0cm}
\begin{array}{cclll}
 & \hspace{0cm}&
[I_{32},I_{43}]_q = I_{42}^+,\  &
[I_{31}^+,I_{43}]_q = I_{41}^+,\ &
[I_{21},I_{42}^+]_q = I_{41}^+, \\
&&
[I_{43},I_{42}^+]_q=I_{32},&
[I_{43},I_{41}^+]_q=I_{31}^+,&
[I_{42}^+,I_{41}^+]_q=I_{21}, \\
&&
[I_{42}^+,I_{32}]_q=I_{43},&
[I_{41}^+,I_{31}^+]_q=I_{43},&
[I_{41}^+,I_{21}]_q=I_{42}^+,
\end{array}                                             \label{O4}
\]
\[
[I_{43},I_{21}]=0,\ \  [I_{32},I_{41}^+]=0\ ,
\]
\[
[I_{42}^+,I_{31}^+]=(q-q^{-1})(I_{21}I_{43}-I_{32}I_{41}^+).   \label{O4p}
\]
%
The first relation in (\ref{FO}) is viewed as definition
for third generator $I_{31}^+$; with this,
the algebra is given in terms of $q$-commutators.
Dual copy of $U'_q({\rm so}_3)$ involves the generator
$I_{31}^-=[I_{21},I_{32}]_{q^{-1}}$ which enters the
relations same as (\ref{FO}), but with $q\to q^{-1}$.
Similar remarks apply to the generators $I_{42}^+$, $I_{41}^+$,
as well as (dual copy of) the whole algebra $U'_q({\rm so}_4)$.

\section{ Deformed Algebras \boldmath$A(n)$ of Nelson and Regge}

For $(2+1)$-dimensional gravity with cosmological constant
$\Lambda<0$, the Lagrangian density involves spin connection
$\omega_{ab}$ and dreibein $e^a$, $\ a,b=0,1,2$, combined in the
${\rm SO}(2,2)$-valued (anti-de Sitter) spin connection
$\omega_{AB}$ of the form
\[   \hspace{2mm}
\omega_{AB} = \left(
\begin{array}{ll}
       \omega_{ab}  &  \frac{1}{\alpha} e^a \\
       -\frac{1}{\alpha} e^b &  0
\end{array}
\right) ,
\]
and is given in the Chern---Simons (CS) form \cite{DJt,Witt}
\[
    \hspace{2mm}
\frac{\alpha}{8} (\hbox{d} \omega^{AB} -
 \frac23 \omega^{A}_{\ F}\wedge \omega^{FB} )\wedge \omega^{CD} \epsilon_{ABCD}.
\]
Here $A,B=0,1,2,3\ $, the metric is $\eta_{AB} = (-1,1,1,-1) $,
and the CS coupling constant is connected with $\Lambda$, so that
$\Lambda = - \frac{1}{3\alpha^2}$. The action is invariant under
${\rm SO}(2,2)$, leads to Poisson brackets and field equations.
Their solutions, i.e. infinitesimal connections, describe
space-time which is locally anti-de Sitter.

To describe global features of space-time, within fixed-time
formulation, of principal importance are the {\it integrated
connections} which provide a mapping $S:$ $\pi_1(\Sigma )\to G$ of
the homotopy group for a space surface $\Sigma$ into the group $G=
SL_+(2,R)\otimes SL_-(2,R)$ (spinorial covering of ${\rm
SO}(2,2)$) and thoroughly studied in   \cite{NRZ}. To generate the
algebra of observables, one takes the traces
\[
c^{\pm}(a)= c^{\pm}(a^{-1}) =\frac12 {\rm tr} [S^{\pm}(a)] \
\]
where
\[
a\in \pi_1, \ \ \ \ S^{\pm }\in SL_{\pm}(2,R).
\]
For $g=1$ (torus) surface $\Sigma$, the algebra of (independent)
quantum observables has been derived   \cite{NRZ}, which turned
out to be isomorphic to the cyclically symmetric Fairlie --
Odesskii algebra      \cite{Od,Far}. This latter algebra, however,
is known to coincide     \cite{GK94} with the special $n=3$ case
of $U'_q({\rm so}_n)$. So, natural question arises whether for
surfaces of higher genera $g\ge 2$, the nonstandard $q$-algebras
$U'_q({\rm so}_n)$ also play a role.

Below, the positive answer to this question is given.

For the topology of spacetime $\Sigma\times{\bf R}$
($\Sigma$ being genus-$g$ surface), the homotopy group
$\pi_1(\Sigma)$ is most efficiently described in terms of
$2g+2=n$ generators $t_1, t_2, \ldots, t_{2g+2}$ introduced in \cite{NR}
and such that
\[
t_1 t_3 \cdots t_{2g+1} = 1,    \hspace{3mm}
t_2 t_4,..., t_{2g+2} = 1,
\hspace{2mm}   \prod_{i=1}^{2g+2} t_i =1.
\]

\noindent
Classical gauge invariant trace elements ($n(n-1)/2$ in total)
defined as
\be
\alpha_{ij} = \frac12 {\rm Tr} (S(t_i t_{i+1} \cdots t_{j-1})), \ \ \ \
                 S\in SL(2,R),
                                \label{clas}
\ee
generate concrete
algebra with Poisson brackets, explicitly found in        \cite{NR}.
At the quantum level, to the algebra with generators
(\ref{clas}) there corresponds quantum commutator algebra $A(n)$
specific for $2+1$ quantum gravity with negative $\Lambda$.
For each quadruple of indices
$\{j,l,k,m\},\ j,l,k,m=1,\ldots,n,$  such that      
\be
i<j<m<k<i \ ,
\ee
\vspace{0.1mm}
\renewcommand{\normalsize}
\noindent
the algebra $A(n)$ of quantum observables reads            \cite{NR}:
\be                                     \label{An}
\begin{array}{clc}
 {} &  [a_{mk}, a_{jl}] = [a_{mj}, a_{kl}] = 0 , &  {}
         \vspace{2mm}     \\
{}  & [a_{jk}, a_{kl}] = (1-\frac1K) (a_{jl} - a_{kl} a_{jk}) , &  {}
\vspace{2mm}      \\
  {}  & [a_{jk}, a_{km}] = (\frac1K -1) (a_{jm} - a_{jk} a_{km}) , &  {}
\vspace{2mm}      \\
  {}  & [a_{jk}, a_{lm}] = (K-\frac1K) (a_{jl} a_{km}- a_{kl} a_{jm}) . &  {}
\end{array}
\ee
Here the parameter $K$ of deformation involves both $\alpha$ and Planck's
constant, namely
\be
   K = \frac{4\alpha - i h}{4\alpha + ih},
    \ \ \alpha^2 = - \frac{1}{3 \Lambda}, \ \  \Lambda < 0.
\ee
Note that in (\ref{clas}) only one copy of the two $SL_{\pm}(2,R)$
is indicated. In conjunction with this, besides the deformed algebra
$A(n)$ derived with, say, $SL_+(2,R)$ taken in (\ref{clas}) and
given by ({\ref{An}}), another identical copy of $A(n)$
(with the only replacement $K\to K^{-1}$) can also be obtained
starting from $SL_-(2,R)$ taken in place of $SL(2,R)$ in (\ref{clas}).
This another copy is independent from the original one:
their generators mutually commute.

\section{ Isomorphism of the Algebras \boldmath$A(n)$ and \boldmath$U'_q({\rm so}_n)$}

To establish isomorphism
\cite{Gnik} between the algebra $A(n)$ from (\ref{An}) and the
nonstandard $q$-deformed algebra $U'_q({\rm so}_n)$ one has to
make the following two steps.

\vspace{3.2mm}
 \ \ --- \  { Redefine}:

    \vspace{-9.2mm}
\[  \hspace{34mm}
       \{ K^{1/2} ({K-1})^{-1} \} a_{ik} \longrightarrow  A_{ik},
\]

 \ \ ---  \ { Identify}:
\vspace{-8.0mm}
\[  \hspace{34mm}
    A_{ik} \longrightarrow  I_{ik},
    \hspace{10mm}
    K \longrightarrow  q .
\]
Then, the Nelson---Regge algebra $A(n)$ is seen to translate
exactly into the nonstandard $q$-deformed algebra $\Uqso$
described above, see (\ref{f2}). We conclude that these two
deformed algebras are isomorphic to each other (of course, for
$K\ne 1$). Recall that $n$ is linked to the genus $g$ as $\
n=2g+2$, while $K=(4\alpha - ih)/(4\alpha + ih)$ with
$\alpha^2=-\frac{1}{\Lambda}$.

Let us remark that it is the bilinear presentation (2) of the
$q$-algebra $U'_q({\rm so}_n)$ which makes possible establishing
of this isomorphism. It should be stressed also that the algebra
$A(n)$ plays the role of ``intermediate'' one: starting with it
and reducing it appropriately, the algebra of quantum observables
(gauge invariant global characteristics) is to be finally
constructed. The role of Casimir operators in this process, as
seen in  \cite{NR}, is of great importance. In this respect let us
mention that the quadratic and higher Casimir elements of the
$q$-algebra $U'_q({\rm so}_n)$, for $q$ being not a root of $1$,
are known in explicit form             \cite{NUW96,GInik} along
with eigenvalues of their corresponding (representation) operators
\cite{GInik}.

As it was shown in detail in \cite{NRZ}, the deformed algebra for
the case of genus $g=1$ surfaces reduces to the desired algebra of
three independent quantum observables which coincides with $A(3)$,
the latter being isomorphic to the Fairlie --- Odesskii algebra
$U'_q({\rm so}_3)$. The case of $g=2$ is significantly more
involved: here one has to derive, starting with the 15-generator
algebra $A(6)$, the necessary algebra of 6 (independent) quantum
observables. J.Nelson and T.Regge have succeeded \cite{NRpr} in
constructing such an algebra. Their construction however is highly
nonunique and, what is more essential, isn't seen to be
efficiently extendable to general situation of $g\geq 3$.

Our goal in this note was to attract attention to the isomorphism
of the deformed algebras $A(n)$ from \cite{NR} and the nonstandard
$q$-deformed algebras $\Uqso$ introduced in \cite{GK91}). The hope
is that, taking into account a significant amount of the already
existing results concerning diverse aspects of $\Uqso$ (the
obtained various classes of irreducible representations [6, 14,
24---28] and others, as well as knowledge of Casimir operators and
their eigenvalues depending on representations, etc.) we may
expect for a further progress concerning construction of the
desired algebra of $6g-6$ independent quantum observables for
space surface of genus $g>2$.

\vskip3mm The research described in this publication was made
possible in part by Award No. UP1-2115 of the U.S. Civilian
Research and Development Foundation (CRDF).

\rezume{%
╟└╤╥╬╤╙┬└══▀ $q$-└╦├┼┴╨ $U'_q({\rm so}_n)$ ─╬ ╩┬└═╥╬┬╬п\\
├╨└┬▓╥└╓▓п: ╧╨╬ $q$-─┼╘╬╨╠╬┬└═╚╔ └═└╦╬├\\ ${\rm SO}(n)$-╤╧▓═╬┬╚╒
╤▓╥╬╩} {╬. ╠. ├ртЁшышъ} {═хёЄрэфрЁЄэ│ $q$-фхЇюЁьютрэ│ рыухсЁш
$U'_q({\rm so}_n)$, чряЁюяюэютрэ│ фхё Є№ Ёюъ│т Єюьє фы  яюЄЁхс
ЄхюЁ│┐ яЁхфёЄртыхэ№, │ёЄюЄэю т│фЁ│чэ ■Є№ё  т│ф ёЄрэфрЁЄэю┐
фхЇюЁьрЎ│┐ ─Ё│эЇхы№фр │ ─ц│ьсю рыухсЁ $U({\rm so}_n)$ │ ьр■Є№
яхЁхф юёЄрээ│ьш трцышт│ яхЁхтруш. ╠ш тштўр║ью ьюцыштх чрёЄюёєтрээ 
$q$-рыухсЁ $U'_q({\rm so}_n)$ є фтюї Ё│чэшї ъюэЄхъёЄрї. ╬фшэ ч эшї
ёЄюёє║Є№ё  $q$-фхЇюЁьєтрээ  $D$-тшь│Ёэю┐ ($D\ge 3$) хтъы│фютю┐
уЁрт│ЄрЎ│┐ эр юёэют│ єчруры№эхээ  ЇюЁьры│чьє ёя│эютшї ё│Єюъ, │э°шщ
фр║ чрёЄюёєтрээ  фю (2+1)-тшь│Ёэю┐ рэЄшфхё│ЄЄхЁ│тё№ъю┐ ътрэЄютю┐
уЁрт│ЄрЎ│┐ (ч Ё│ьрэютю■ яютхЁїэх■ Ёюфє $g$) є я│фїюф│ ═хы№ёюэр │
╨хфцх.}

\rezume{%
╧╨╚╠┼═┼═╚┼ $q$-└╦├┼┴╨ $U'_q({\rm so}_n)$ ┬ ╩┬└═╥╬┬╬╔\\ ├╨└┬╚╥└╓╚╚:
╬ $q$-─┼╘╬╨╠╚╨╬┬└══╬╠\\ └═└╦╬├┼ ${\rm SO}(n)$-╤╧╚═╬┬█╒ ╤┼╥┼╔} {└.
╠. ├ртЁшышъ} {═хёЄрэфрЁЄэ√х $q$-фхЇюЁьшЁютрээ√х рыухсЁ√ $U'_q({\rm
so}_n)$, яЁхфыюцхээ√х фхё Є№ ыхЄ Єюьє эрчрф т ёт чш ё
яюЄЁхсэюёЄ ьш ЄхюЁшш яЁхфёЄртыхэшщ, ёє∙хёЄтхээю юЄышўр■Єё  юЄ
ёЄрэфрЁЄэющ фхЇюЁьрЎшш ─ЁшэЇхы№фр ш ─цшьсю рыухсЁ $U({\rm so}_n)$
ш юсырфр■Є Ё фюь яЁхшьє∙хёЄт. ╠√ шчєўрхь тючьюцэюх яЁшьхэхэшх
$q$-рыухсЁ $U'_q({\rm so}_n)$ т фтєї Ёрчышўэ√ї ъюэЄхъёЄрї. ╬фшэ шч
эшї ърёрхЄё  $q$-фхЇюЁьшЁютрэш  $D$-ьхЁэющ ($D\ge 3$) ¤тъышфютющ
уЁртшЄрЎшш эр юёэютх юсюс∙хэш  ЇюЁьрышчьр ёяшэют√ї ёхЄхщ, фЁєующ
--- яЁшьхэхэш  ъ (2+1)-ьхЁэющ рэЄшфхёшЄЄхЁютёъющ ътрэЄютющ
уЁртшЄрЎшш (ё Ёшьрэютющ яютхЁїэюёЄ№■ Ёюфр $g$) т яюфїюфх ═хы№ёюэр
ш ╨хфцх.}

\end{multicols}
\end{document}